\documentclass[twocolumn,pre,nofootinbib,showpacs,superscriptaddress]{revtex4-1}

\newcommand{\beq}{\begin{eqnarray}}
\newcommand{\eeq}{\end{eqnarray}}

\usepackage{graphicx,amsmath,tabularx}

\begin{document}

\title{Non-equilibrium evolution of window overlaps in spin glasses}
\author{Markus Manssen}
\affiliation{Institute of Physics, Carl von Ossietzky University, Oldenburg,
Germany}
\author{Alexander K. Hartmann}
\affiliation{Institute of Physics, Carl von Ossietzky University, Oldenburg,
Germany}
\author{A.~P.~Young}
\affiliation{Department of Physics, University of California, Santa Cruz,
California 95064, USA}

\date{\today}

\begin{abstract}
We investigate numerically the time dependence of ``window" overlaps in a 
three-dimensional Ising spin
glass below its transition temperature after a rapid quench. 
Using an efficient GPU implementation, we are able to study large
systems up to lateral length $L=128$ and up to long times of $t=10^8$
sweeps.
We find that the
data scales according to the ratio of the window size $W$ to the
non-equilibrium coherence length $\xi(t)$.  We also show a substantial change
in behavior if the system is run for long enough that it globally
equilibrates, i.e.\ $\xi(t) \approx L/2$, where $L$ is the
lattice size. This indicates that the local
behavior of a spin glass depends on the spin configurations (and presumably
also the bonds) far away. We compare with similar simulations
for the Ising ferromagnet. Based on these results,
we speculate on a connection between the
non-equilibrium dynamics discussed here and averages computed theoretically
using the ``metastate". 
\end{abstract}

\maketitle

\section{Introduction}
\label{sec:intro}

Spin glasses \cite{binder:86,mezard:87,young:98} 
below their transition temperature are not in equilibrium, except for very
small sizes in some simulations. One therefore needs to be able to describe
non-equilibrium behavior, and a lot of attention
numerically~\cite{rieger:93,kisker:96,marinari:96,yoshino:02,bellettietal:09,manssen:14}
has been focussed
on the evolution of the system after a rapid quench to temperature $T$ below
the transition temperature $T_c$.  Locally, spins establish correlations so
one anticipates that they will be correlated up to some distance, the
coherence length $\xi(t)$, which slowly increases with time. For distances
longer than $\xi(t)$ correlations will decay exponentially, while at shorter
distances they will decay more slowly than that. Empirically
one
finds~\cite{rieger:93,kisker:96,marinari:96,yoshino:02,bellettietal:09,manssen:14}
that the growth of $\xi(t)$ is compatible with a small power of $t$
(although a logarithmic growth cannot be fully excluded using the 
available data), 
written as $\xi(t) \sim
t^{1/z(T)}$ where, $z(T)$, a non-equilibrium dynamic exponent is found to
depend on the ratio $T/T_c$.

To understand the nature of the spin glass state one needs
local probes, see e.g.~\cite{newman:03} and references therein.
A useful local probe is the distribution
of the overlap $q$ of the spins in
two copies of the system in a window of linear size $W$. 
Equilibrium properties of window overlaps have
been studied numerically before~\cite{marinari:98e}, but here we focus on 
their non-equilibrium behavior, which has not received much attention before
apart from Ref.~\cite{marinari:96} which studied the non-equilibrium evolution
of a dimensionless ratio of cumulant averages evaluated in windows of
different size.
In this
paper we study the time dependence of the window overlap distribution
$P_W(q)$, in a non-equilibrium
situation. We find that the distribution scales as a function of the ratio of
the window size $W$ to $\xi(t)$. The non-equilibrium window
overlap distribution is very
different from the global equilibrium overlap distribution $P(q)$
in the mean field
theory of Parisi~\cite{parisi:79,parisi:80,parisi:83}. In particular $P_W(0)$
depends quite strongly on $W$.
However, if the system is run for a time long
enough for the system to globally equilibrate, i.e. $\xi(t) \simeq L/2$, then
we find a
change in the form of $P_W(q)$, which happens rapidly
when viewed on a logarithmic time scale,
such that $P(0)$ then has a rather a weak
dependence on $W$ and is quite similar to the $q=0$ value of
Parisi's global
equilibrium overlap distribution~\cite{parisi:83} $P(q)$.
The strong change in behavior when $\xi(t) \simeq L/2$
indicates that local spin correlations are sensitive to spin orientations, and
presumably also to the values of the interactions, at large (or at
least intermediate) distances. 

The theoretical description of spin glasses below $T_c$ is complicated. One
approach developed in recent years is known as the
``metastate"~\cite{newman:96b,newman:97,aizenman:90}.
In this paper
we also speculate on a possible connection between non-equilibrium
correlations following a quench, and averages computed according to the
metastate.

The plan of this paper is as follows. In Sec.~\ref{sec:metastate} we describe
the metastate and a possible connection between quantities calculated from
it and non-equilibrium averages following a quench. The model we simulate and
the quantities we calculate are described in Sec.~\ref{sec:model}. The results
of the simulations are given in Sec.~\ref{sec:results},
together with corresponding results for a pure Ising
ferromagnet, and our conclusions
summarized in Sec.~\ref{sec:concl}.

\section{Averaging in spin glasses; the Metastate and Dynamics}
\label{sec:metastate}

In systems undergoing phase transitions it is desirable to know what are the
various possible states to which the system can evolve
below the transition temperature
$T_c$. A simple example is the Ising ferromagnet in zero magnetic field
for which there is just a
pair of states below $T_c$, related by flipping all the spins, the ``up''and
``down'' spin states. If the system is in one of these states then
``connected'' correlation functions vanish at large distances, e.g.
\begin{equation}
\lim_{|{\bf R}_i - {\bf R}_j| \to \infty}  \left[\,
\langle S_i S_j \rangle - \langle S_i \rangle \langle S_j \rangle
\, \right]= 0\, ,
\label{clust}
\end{equation}
which is known as ``clustering'' of the correlation functions. The angular
brackets $\langle \cdots \rangle$ denote a thermal average,
here restricted to one of these states to capture the
symmetry breaking.  By contrast if
we simply perform the Boltzmann sum we give equal weight to both of these states, the
symmetry is not broken so $\langle S_i \rangle = \langle S_j\rangle = 0$,
and hence the
two terms in Eq.~\eqref{clust} do not cancel at large distances and the
correlation functions do not have a clustering property. States which do not
have a clustering property are called ``mixed'' states and those that do,
like the ``up'' and ``down'' states of the ferromagnet, are called ``pure" states.

To keep the description of the system as simple as possible it is 
desirable to use clustering (i.e.\ pure) states.
In many cases
this is easy because they are just the
different states in which a global symmetry of the Hamiltonian is broken.
However, in more complicated situations such as spin glasses, there can be
pure states not related by any symmetry and so characterizing them can be
quite difficult. Nonetheless, it is
argued~\cite{fisher:86,fisher:87b,fisher:87,fisher:88,newman:92,newman:96b,newman:97,newman:03,read:14}
to be important to describe spin glasses in terms of pure states rather than
by computing the Boltzmann sum. The latter is done, for example, in the
Parisi~\cite{parisi:79,parisi:80,parisi:83}
solution of the infinite-range Sherrington-Kirkpatrick
(SK)~\cite{sherrington:75} model.

\begin{figure}[!tb]
\begin{center}
\includegraphics[width=0.9\columnwidth]{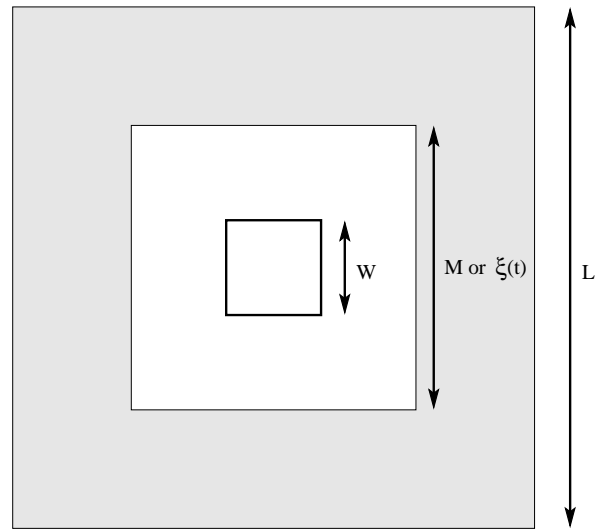}
\caption{
The length scales that are needed to discuss the Aizenman-Wehr (AW) metastate.
The overall size of the system is $L$, which is assumed to be very large
and has periodic or free boundary
conditions. We consider an outer region, of size
between $M$ and $L$, shown shaded, where we average over different bond
configurations, and an inner region, unshaded, where we consider just a single
set of bonds. Spin correlations will be studied in a window of size
$W$, less than $M$. In the metastate, we require that
the different length all ultimately tend to infinity such that
$L \gg M \gg W \gg 1$. In our simulations, the length scales are, of course,
finite (actually quite small) but we shall still view the situation in the 
simulations as analogous to the
theoretical discussion, in which, as for the metastate,
$L$ is the size of the system (with periodic
boundary conditions) and $W$ is a small
region where we measure correlations, but now
$M$, the intermediate scale, is the \textit{non-equilibrium coherence
length} $\xi(t)$, the scale to which correlations have developed after a quench
at time $t=0$. In the simulations we can run for long enough, and
take sufficiently small window sizes, to get data in the region where $\xi(t)
> W$ as shown. 
\label{fig:LMW}}
\end{center}
\end{figure}

To define pure states in general, consider the situation in
Fig.~\ref{fig:LMW}. The overall system is of very large size $L$ and has free
or periodic boundary conditions. We compute the thermal average exactly, and
determine the correlation functions in a much smaller window of size
$W$, somewhere in the bulk of the system. These correlations may have the
clustering property, in which case the window is in a pure state, or they may
not in which case it is in a mixed state. In fact, since we consider only zero
field, states come in symmetry-related pairs, so the simplest situation would
be a single pair of pure states. 

But for a system like a spin glass, the correlations in the
window could depend sensitively on the choice of interactions in distant
regions of the of the system,
perhaps even in a chaotic manner, an aspect first pointed out
explicitly by Newman and Stein (NS)~\cite{newman:92}.
To investigate this we divide the system
of size $L$ into an inner region of size $M$, larger than the window of size
$W$ which is in the middle of it, and an outer region between $L$ and $M$. We
then change the bonds in the outer region and recompute the correlation
functions in the window. Eventually we let all sizes tend to infinity with
$L \gg M \gg W \gg 1$. It is 
possible that the state of the window is always the same as
one changes the bonds in the outer region.
However, it is also possible that the
state changes, perhaps chaotically, as one changes the bonds in the outer
region. Several possible situations have been discussed in detail:
\begin{itemize}
\item
For each set of bonds in the outer region one has only a single pair of pure
states, and one finds the \textit{same} pair for every set of outer bonds. 
This is called the ``droplet model'' the theory for which has been developed
in the greatest detail by Fisher and
Huse~\cite{fisher:86,fisher:87b,fisher:87,fisher:88}.
\item
For each set of bonds in the outer region one has only a single pair of pure
states, but this pair varies chaotically as one changes the outer bonds. This
is the ``chaotic pairs" picture of NS~\cite{newman:92}.
\item
For each set of bonds in the outer region one has a mixed state, and this
mixes changes in a chaotic way as the outer bonds are changed.
This is called the``replica symmetry breaking''
(RSB) picture\footnote{NS~\cite{newman:96b,newman:97}
call this the ``non-standard'' RSB
picture, because they showed that a different, ``standard'', RSB picture is
not viable. As also emphasized recently by Read~\cite{read:14}, the
``non-standard'' picture is the only viable RSB picture, so we shall omit the
term ``non-standard'' and just refer to this scenario as the ``RSB picture".
In fact, Read also shows that the RSB calculations lead \textit{directly} to
the ``non-standard'' picture.}
since it is the generalization to finite-range models of
Parisi's~\cite{parisi:79,parisi:80,parisi:83}
solution of the infinite-range SK model. The name arises because
Parisi's original solution used
the replica method to average over the disorder.
\end{itemize}

In order to describe the states of a spin glass one needs to give a
\textit{statistical} description of the different states the window can be in
as the bonds in the outer region are varied. NS~\cite{newman:96b,newman:97}
call this
the metastate. The description that we give here is actually a little
different from that of NS and is due to Aizenman and Wehr (AW)~\cite{aizenman:90}. In
NS's approach there is no intermediate scale $M$ and one looks
at the correlations in the window as the system size $L$ is grown leaving the
bonds already present unchanged. It is expected~\cite{newman:97b}
that the two forms of the
metastate are equivalent. In agreement with Read~\cite{read:14}
we find that it is easier to discuss
the AW metastate.

The AW metastate average is therefore performed by first
doing a thermal average for the whole system,
denoted by $\langle \cdots \rangle$, followed by an average over the bonds in
the outer region, denoted by $[ \cdots ]_\text{out}$. Following
Read~\cite{read:14} we call this the
metastate-averaged state (MAS). Hence, if $i$ and $j$ lie within the window,
the spin glass correlation function of their spins in the MAS is given by
\begin{equation}
C_{i j} = \left[\, \langle S_i S_j \rangle \, \right]^2_\text{out} \, ,
\end{equation}
(note the location of the square).
After this average is done
one can also average over the bonds in the inner region,
which we denote by $[ \cdots ]_\text{in}$.
We will present
data for the window overlap distribution for which averaging over the bonds in
the inner region is, strictly, speaking, 
unnecessary since
translation invariant
MAS averages are self-averaging~\cite{newman:96,newman:96b,newman:97}.
However, in practice, this last average \textit{is} done in 
simulations to improve statistics.

It is interesting to ask how the MAS average $C_{ij}$ varies at large
distance $R_{ij} \ (\equiv |{\bf R}_i - {\bf R}_j|)$ according to the three
scenarios mentioned above:
\begin{itemize}
\item
In the droplet picture one finds always the same pair of thermodynamic states
so presumably
\begin{equation}
\lim_{R_{ij}\to\infty} C_{ij} = q^2 \, ,
\label{Cij_droplet}
\end{equation}
where $q = \langle S_i \rangle^2$ is the Edwards-Anderson order parameter,
which is well defined if we add a small symmetry-breaking field to remove the
degeneracy between the pair of pure states. Equation \eqref{Cij_droplet} then
follows because of clustering of correlations in a single pure state, see
Eq.~\eqref{clust}. We should mention, though, that the approach to the
constant value of $q^2$ is expected to be quite slow, a power-law rather than an
exponential, and so, for the values of $R_{ij}$ that one can simulate, one may be 
far from the constant value (David Huse, private communication).
\item
In the chaotic pairs picture, correlations in the window alter, in sign as
well as magnitude, as the outer bonds are varied. Hence, according to
Read~\cite{read:14}, it is expected 
that $C_{ij}$ tends to zero, presumably as a power law, which is
commonly written as
\begin{equation}
C_{ij}  \propto {1 \over R_{ij}^{d-\zeta}} \, ,
\label{zeta}
\end{equation}
for $R_{ij} \to\infty$, which defines the exponent $\zeta$. 
\item
In the RSB picture, which also has many states, MAS averaged correlations are similarly
expected to decay as the power law in Eq.~\eqref{zeta}. In fact $\zeta$ has
been calculated in mean field
theory~\cite{dedominicis:98,dedominicis:06,marinari:00a,read:14}
(corresponding to $d >6$) assuming RSB,
with the result $\zeta = 4$.
\end{itemize}

A large spin glass system is not in thermal
equilibrium below $T_c$.
Results from the Boltzmann sum do not, therefore, correspond to
experimental observations which are inevitably in a non-equilibrium situation.
Are MAS averages any better in this regard?
It is tempting to think
so for the following reason. 

Imagine quenching the spin glass to below $T_c$ and observing correlations in
a local window of size $W$. Correlations will develop up to some
coherence length $\xi(t)$ which grows slowly with time.
How does one expect the non-equilibrium correlation function
\begin{equation}
C_t(i, j) = [\langle S_i(t) S_j(t) \rangle^2] \, ,
\label{Ct}
\end{equation}
where $[\cdots]$ denotes an average over \textit{all} the bonds, 
to vary as a function of $R_{ij}$?
Let us assume that time is large enough that $\xi(t) > W$.
We postulate that thermal fluctuations of the
spins outside the window
at a distance $\xi(t)$ and greater,
which are not equilibrated with respect to spins
in the window, 
effectively generate a random
noise to the spins in the window which plays a similar role to the random
perturbation coming from changing the bonds in the outer region according to
the AW metastate, see Fig.~\ref{fig:LMW}. Thus we suggest that
$\xi(t)$ is analogous to
the intermediate scale $M$, separating inside and outside regions, in the
construction of the metastate. This is indicated in Fig~\ref{fig:LMW}.
After
this work was submitted it was brought to our attention that a similar picture of
non-equilibrium dynamics following a quench was discussed earlier by White
and Fisher~\cite{white:06}. They denote the state obtained after a quench as
the ``maturation metastate'' and
the distribution of states in the AW or NS
picture as the ``equilibrium metastate''. Here we speculate that these might
be the same. We thank Nick Read for bringing this paper to our attention.

This analogy suggests that the decay of correlations determined from the
metastate may be the same as the decay of correlations following a quench, on
scales shorter than the coherence length. We note that NS have also discussed
dynamics following a quench~\cite{newman:07b,newman:99b} from a rigorous
point of view.

There have been many simulations which investigate the time dependence of 
correlations following a
quench~\cite{rieger:93,kisker:96,marinari:96,yoshino:02,bellettietal:09,manssen:14}.
Interestingly these papers \textit{do}
see a power law decay of the correlation function
in Eq.~\eqref{Ct} for sufficiently long times that $\xi(t) > R_{ij}$,
i.e.
\begin{equation}
C_t(i, j) \propto {1 \over R_{ij}^{\kern1pt\raisebox{1pt}{\scriptsize$\alpha$}}}
\quad \text{for} \quad R_{ij} \ll \xi(t) \ll L \, .
\label{Ct_alpha}
\end{equation}
The exponent $\alpha$ is found to be
about $1/2$ in three
dimensions~\cite{rieger:93,kisker:96,marinari:96,yoshino:02,bellettietal:09,manssen:14}.
Equation \eqref{Ct_alpha} is of the same form as Eq.~\eqref{zeta}
which is obtained from metastate
calculations for the
Edwards-Anderson model~\cite{marinari:00a,read:14} in the mean field
approximation,
assuming the RSB picture. The droplet theory predicts a different result,
namely Eq.~\eqref{Cij_droplet}, 
though, of course, the numerical data may not be at large enough length
scales to be in the asymptotic scaling regime. 

\section{The model and quantities to be calculated}
\label{sec:model}

We simulate the Edwards-Anderson~\cite{edwards:75} Ising spin glass model with
Hamiltonian
\begin{equation}
\mathcal{H} = - \sum_{\langle i, j\rangle} J_{ij} S_i S_j \, ,
\end{equation}
where the spins $S_i$ take values $\pm 1$ and are on the sites of a simple
cubic lattice with $N = L^3$ spins with periodic boundary conditions. The
quenched interactions $J_{ij}$ are between nearest neighbors and take values
$\pm 1$ with equal probability. The latest determination of the
transition temperature of this model
is $T_c= 1.102(3)$~\cite{baity-jesietal:13}.
Here we work at a fixed temperature of
$T= 0.8 \simeq 0.73 T_c$.
Most of the simulations are for system size $L=128$, which can not
be brought to equilibrium in available computer time, but we also
perform some simulations at smaller sizes to investigate the change in behavior
when the system reaches global equilibrium. The number of samples simulated
for each size is shown in Table~\ref{tab:samps}.

\begin{table}
\begin{tabular}{|r|r|r|}
\hline
 & \multicolumn{2}{c|}{$N_\text{samp}$} \\
 \multicolumn{1}{|c|}{$L$} & Spin glass & Ferromagnet \\
\hline \hline
 128 & 192 & 64\\
 64 & - & 512\\
 32 & - & 512\\
 20 & 512 & -\\
 16 & 768 & 2048\\
 12 & 1024 & -\\
\hline
\end{tabular}
\caption{
The number of samples studied for different system sizes.
}
\label{tab:samps}
\end{table}

We run two copies of the system with the same bonds but different initial 
random spin configurations, which we
quench to $T=0.8$ at time $t = 0$,
and then let the system evolve. To perform long runs on large
lattices we have implemented an efficient Monte Carlo code on GPUs,
see~\cite{manssen:14} for details. At a logarithmically increasing set of
times we store the spin configurations from which we calculate the correlation
function in Eq.~\eqref{Ct} as a function of $R_{ij}$ at different times. 

We also compute the time-dependent window overlap distribution defined by
\begin{equation}
P_W(q) = \left[\langle \delta \left(q - q^{1,2}\right)\rangle \right] \, , 
\end{equation}
where $q^{1,2}$, the window overlap between replicas ``$(1)$'' and ``$(2)$'', is
\begin{equation}
q^{1,2} = {1 \over W^d} \, \sum_{i=1}^{W^d} S_i^{(1)} S_i^{(2)} \, ,
\end{equation}
in which the sum is over the sites in the window
and, for ease of notation,
we have suppressed an index $t$ on $P_W(q)$ which would indicate that it
also depends on time.
To improve statistics we average over all non-overlapping
windows of size $W$. The number of these is
$\left\lfloor\, L/W\, \right\rfloor^d$ where
$\lfloor \cdots \rfloor$ indicates rounding down to the
nearest integer. In addition, we smooth the data by computing, for each discrete
value of the overlap, $q_0$ say,
an average of the distribution on neighboring $q$-values weighted by a 
normalized kernel
which falls to zero as $|q-q_0|$ increases~\cite{banosetal:10}. 

\section{Results}
\label{sec:results}

\subsection{Spin Glass}
\label{sec:results_sg}

\begin{figure}[!tb]
\begin{center}
\includegraphics[width=\columnwidth]{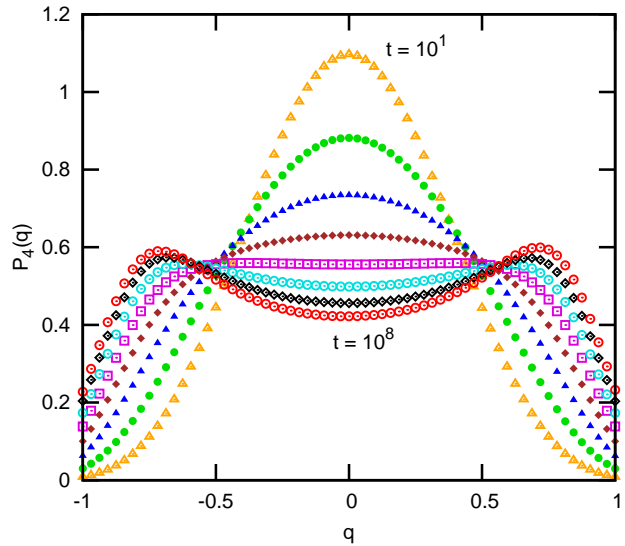}
\caption{(Color online)
A representative set of results for the window overlap distribution, for window
size $W=4$ for lattice size $L=128$ at $T = 0.8$. Data is shown for times
$t=10^k$, where $k = 1, 2, \cdots, 8$.
\label{fig:P_4_t}
}

\end{center}
\end{figure}
\begin{figure}[!tb]
\begin{center}
\includegraphics[width=\columnwidth]{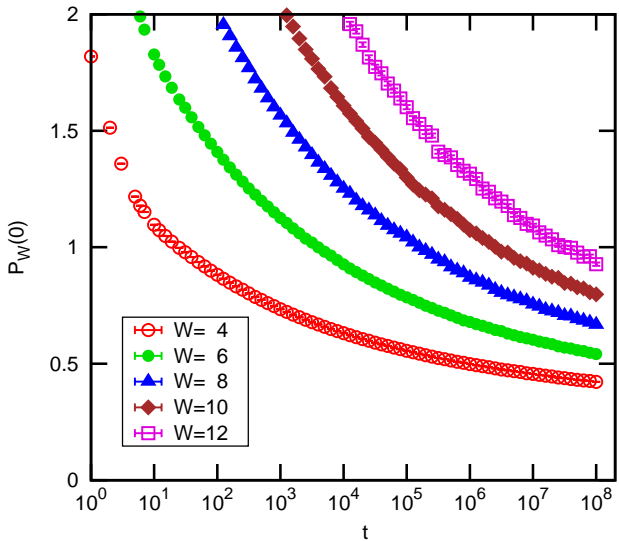}
\caption{(Color online)
Results for the weight in the window overlap distribution
at $q=0$ for different
window sizes as a function of time. The lattice size is $L=128$ and $T=0.8$.
\label{fig:P_0_t}
}
\end{center}
\end{figure}

An example of our data for the window overlap distribution is shown in
Fig.~\ref{fig:P_4_t} for $W=4$.
One sees an evolution from a single peak structure at
short times, presumably Gaussian, to a two-peaked structure at long times. 
For larger window sizes,
the distribution evolves more slowly, as shown in
the data for $P_W(0)$, the weight of the distribution at $q=0$, for
different sizes in
Fig.~\ref{fig:P_0_t}.

\begin{figure}
\begin{center}
\includegraphics[width=0.9\columnwidth]{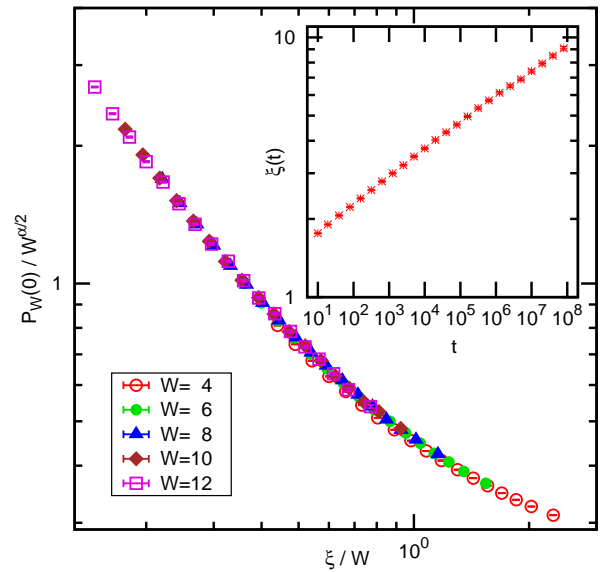}
\caption{(Color online)
 The main figure shows a scaling plot of
$P_W(0)/W^{\alpha/2}$ against $\xi/W$, in which
we used the results for $\xi(t)$
shown in the inset, which are obtained in Ref.~\cite{manssen:14},
and took the value 
$\alpha = 0.438$ also from Ref.~\cite{manssen:14}.
The data
collapse is excellent. The data is for system size $L = 128$ and $T=0.8$.
\label{fig:P_B_scale}
}
\end{center}
\end{figure}

We would like to perform a scaling collapse of
the data in Fig.~\ref{fig:P_0_t} to ascertain the
dependence of $P_W(0)$ 
on $t$ and $W$. However, rather than scaling with respect to $t$ we find it
better to scale with the dynamic coherence length $\xi(t)$.
At long times, where $\xi(t) \gg R_{ij}$, the
time-dependent correlation function in Eq.~\eqref{Ct} varies with 
an inverse power of $R_{ij}$ as shown in Eq.~\eqref{Ct_alpha}, so a natural scaling
ansatz is
\begin{equation}
C_t(i, j) = {1 \over R_{ij}^{\kern1pt\raisebox{1pt}{\scriptsize$\alpha$}}}\, g\left({R_{ij} \over \xi(t)} \right).
\label{Ct_scale}
\end{equation}
The coherence length
$\xi(t)$ can be taken from a ratio of moments of
$C_t(i,j)$~\cite{bellettietal:09}, e.g.
\begin{equation}
\xi(t) = {\int_0^{L/2} r^2\, C_t(r) \, d r \over
\int_0^{L/2} r\, C_t(r) \, d r }\, .
\label{moment}
\end{equation}
In practice the integral is performed along $x, y$ and $z$ axes. The data for
$\xi(t)$ obtained in this way in Ref.~\cite{manssen:14}
is shown in the inset to Fig.~\ref{fig:P_B_scale}.

Note that this calculation of $\xi(t)$ did not make any reference to a
window. However, if we compute the second moment of the window overlap
distribution,
$\left[\langle q^2 \rangle\right]$, we note first that it is just the average
of $C_t(i, j)$ over all sites $i$ and $j$ in the window since
\begin{align}
[\langle q^2 \rangle] 
&= \frac 1 {W^6} \sum_{i, j} \left[\, \left\langle S_i^{(1)}S_i^{(2)}
S_j^{(1)}S_j^{(2)}\right\rangle\, \right] \nonumber \\
&= 
\frac 1 {W^6} \sum_{i,j} \left[\, \left\langle  S_i^{(1)}
S_j^{(1)}\right\rangle \left\langle S_i^{(2)} S_j^{(2)}\right\rangle\, \right] \nonumber \\
&= {1 \over W^6} \, \sum_{i, j} C_t(i, j) \, .
\end{align}
Using Eq.~\eqref{Ct_scale} this can be written as
\begin{align}
[\langle q^2 \rangle] 
&=
\frac 1 {W^6} \int_{R,R'} dR\,dR' {|R-R'|^{-\alpha}} g\left(|R-R'|/\xi\right)
\nonumber \\
& \sim \frac 1 {W^6} \int_R \int_0^{W/2}dr\ 4\pi r^2  {r^{-\alpha}}
g\left(r/\xi\right) \nonumber \\
&= 
W^{-\alpha} 4 \pi \int_0^{1/2}dx x^{2-\alpha} g(xW/\xi) \nonumber \\
&= W^{-\alpha} f\left({W \over \xi}\right) \, ,
\label{scale_q2}
\end{align}
where we used the substitution $x = r/W$ in the next to last line.


\begin{figure}
\begin{center}
\includegraphics[width=\columnwidth]{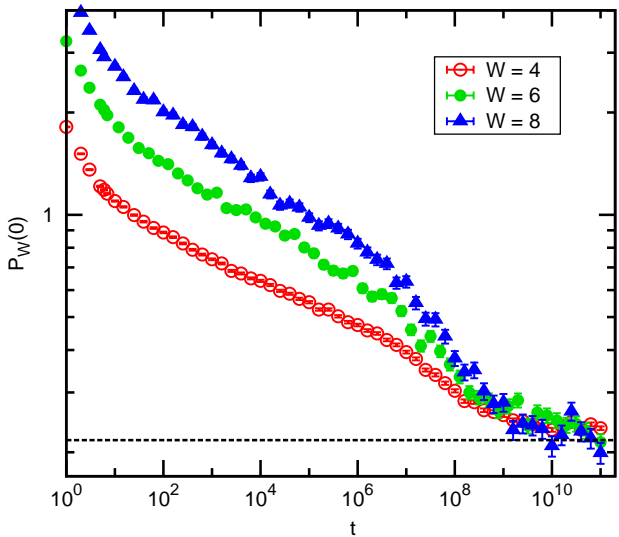}
\caption{(Color online)
Data for $P_W(0)$ for different values of $W$ for size $L=16$. At time in the
range $10^7$--$10^9$ this rather small system fully equilibrates leading to a
decrease in the data which is quite rapid on this log
scale. The dashed line shows the equilibrium value
of the \textit{bulk} order parameter distribution for $L=16$, i.e.\ $W = L =
16$. One sees that the \textit{equilibrium}
values of $P_W(0)$ for $W < L$ are very
similar to that of the bulk overlap.
\label{fig:P_0_L16_t}
}
\end{center}
\end{figure}

\begin{figure}
\begin{center}
\includegraphics[width=0.9\columnwidth]{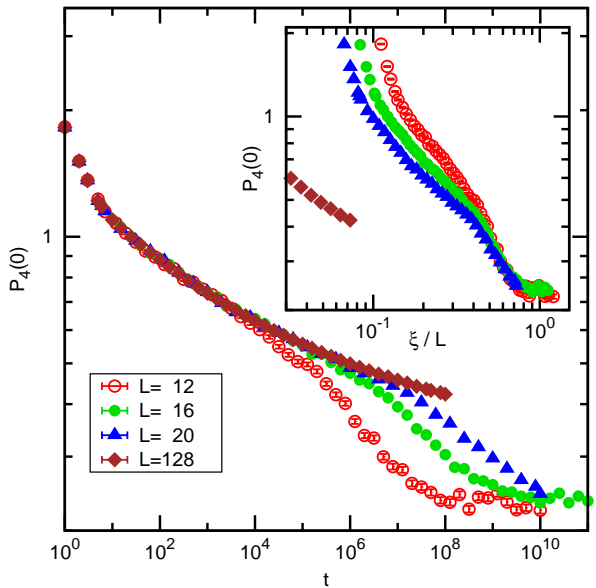}
\caption{(Color online)
Data for $P_4(0)$ for different system sizes at $T=0.8$. For the three smaller sizes,
$L=12,16$ and $20$ the system equilibrates, leading to a pronounced drop in
the data, at a time which increases with $L$. The inset plots the same data
against $\xi/L$ where for $\xi$ we use a fit to the data in the inset of
Fig.~\ref{fig:P_B_scale}. The collapse of the
data in the region where it decreases, which is quite rapid on 
this log scale, shows that the decrease occurs
when $\xi$ is a fixed fraction (about $1/2$) of the system size, indicating
full equilibration.
\label{fig:P_4_0_L}
}
\end{center}
\end{figure}

If we divide $q$ by an arbitrary scale factor $\lambda$ the distribution of
$q' \, (= q / \lambda)$ is $\overline{P}(q')$ where
$P(q) = \lambda^{-1} \overline{P}(q / \lambda)$ because both distributions are
normalized. If 
we take $\lambda = \sigma$, the standard deviation of $P(q)$,
then $P(0) = \sigma^{-1} \overline{P}(0)$. But
$\overline{P}(q')$ has
standard deviation unity, and so, if the distribution is smooth and extends
down to the origin, we have $\overline{P}(0) \sim 1$ and hence $P(0) \sim
\sigma^{-1}$.
Consequently, from Eq.~\eqref{scale_q2}, the expected scaling of $P_W(0)$ is
\begin{equation}
P_W(0) = W^{\alpha/2} \, F\left({\xi \over W}\right) \, .
\label{PW0}
\end{equation}
For $t$ large but still smaller than the time to equilibrate the whole
system, 
the dependence on $\xi$ must drop out and so
\begin{equation}
P_W(0) \propto W^{\alpha / 2 } \quad \text{for} \quad W \ll \xi(t) \ll L \, .
\end{equation}
For short times where $\xi(t) \ll W$ the spins in the window are random, so
the mean square window overlap goes like $1/W^d$ (in $d$ dimensions) and 
consequently
$P_W(0) \propto W^{d/2}$.  Presumably we then have $F(x) \propto
x^{-(d-\alpha)/2}$ for $x \to 0$. This actually gives $P_W(0) \propto W^{d/2} /
\xi^{(d-\alpha)/2}$ but when $\xi(t) \lesssim 1$ corrections to scaling occur which
cause $\xi$ to be replaced by a cutoff of order unity and so one obtains the
desired result.

We take $\xi(t)$ from Ref.~\cite{manssen:14}, evaluated according to
Eq.~\eqref{moment}, and also use the value of $\alpha$ from
Ref.~\cite{manssen:14}, $\alpha = 0.438(11)$.
This exponent has also been
computed in Ref.~\cite{bellettietal:09} with a very similar value, $\alpha = 0.442(11)$.
The result of scaling the data in Fig.~\ref{fig:P_0_t} according to
Eq.~\eqref{PW0} is shown in the main part of Fig.~\ref{fig:P_B_scale}. Clearly
the scaling collapse works very well.

The power law decay of correlation in Eq.~\eqref{Ct_alpha}, and the resulting behavior
of the window order parameter distribution in Eq.~\eqref{PW0}, are for a
non-equilibrium situation where $\xi(t) \ll L$. However, we shall
now see that a dramatic change
occurs at sufficiently long times that global equilibrium occurs, i.e.\ when
$\xi(t) \sim L/2$. In this region, we
will find that the correlation function no longer decays to
zero because there is spin glass order in equilibrium, and the weight of the
window distribution at $q=0$~\cite{marinari:98e,marinari:00a} becomes roughly independent
of window size rather than increasing with window size in the manner shown in
Eq.~\eqref{PW0}.

We demonstrate this change in behavior for the window overlaps
explicitly as a function of time in Figs.~\ref{fig:P_0_L16_t}
and \ref{fig:P_4_0_L}. Since
equilibrating size $L=128$ is completely infeasible
we show data for
smaller sizes which we \textit{can} bring to global equilibrium.  Figure
\ref{fig:P_0_L16_t} shows results for $L=16$ with window sizes $W=4, 6$ and 8.
A rapid decrease is seen for $t$ in the region $10^7$--$10^9$ to a value which
is independent of window size. As will be confirmed in Fig.~\ref{fig:P_4_0_L},
the data after the drop represents global equilibrium.  The dashed line in
Fig.~\ref{fig:P_0_L16_t} is the \textit{bulk} value of the equilibrium overlap
distribution at $q=0$ and we see that this value is very similar to that of
equilibrium \textit{window} overlaps, as was also found
earlier~\cite{marinari:98e,marinari:00a}.

To confirm that this change in behavior occurs when $\xi \sim L / 2 $
we plot results
for $P_W(0)$ for a fixed window size but different system sizes in
Fig.~\ref{fig:P_4_0_L}. For short times the data is independent of $L$
indicating that $\xi \ll L$, but at later times a more rapid decrease occurs at
a time which increases with $L$. The inset shows the data plotted against
$\xi/L$ clearly demonstrating that the region with rapid decrease occurs when
$\xi/L$ is about $1/2$. This confirms that the decrease is associated with
complete equilibration of the system. 



\subsection{Ferromagnet}
\label{sec:results_f}

For comparison we also did simulations of the ferromagnet,
$p=1.0$, at temperature $T=3.6$. Since $T_c \simeq 4.51$ for the
ferromagnet, this corresponds to
$T=0.80 T_c$, a similar fraction of $T_c$
as used in the spin glass simulations. 
The number of samples is detailed in
Table~\ref{tab:samps}.
It should be pointed out
that we are still using a single random number for multiple samples (due to
multi-spin coding techniques) but with different initial configurations, as is
common practice for spin-glass simulations. But in ferromagnetic
equilibrium this causes the samples to become almost completely
correlated. However, before equilibration, due to the different
initial configurations, the dynamics of the different samples is different.

\begin{figure}
\begin{center}
\includegraphics[width=\columnwidth]{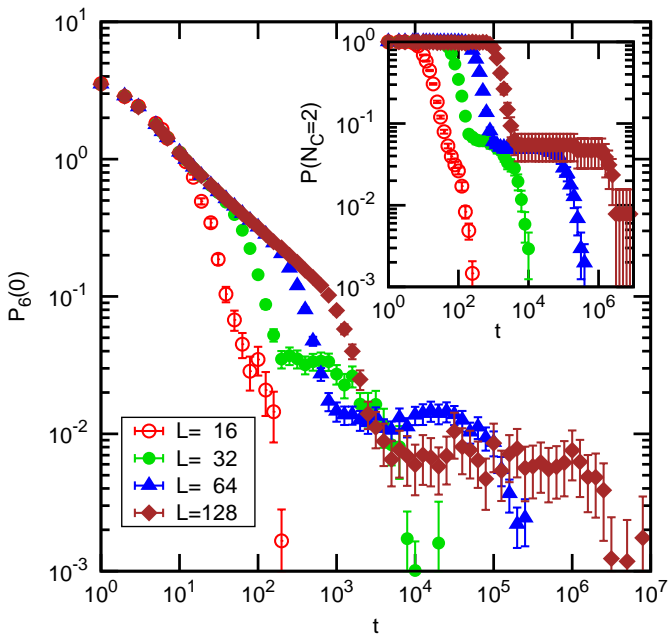}
\caption{(Color online)
Data for $P_6(0)$ for different system sizes at
$T=3.6 \simeq 0.80 T_c$ for the ferromagnet. 
The plateau at intermediate times becomes longer for larger $L$.
The explanation is that
a fraction of the runs gets stuck in a state with 
two big domains for a long time, thus contributing a certain number 
of overlaps with $q \simeq 0$. The inset shows the probability that
two large domains coexist.
\label{fig:P60a_inset}
}
\end{center}
\end{figure}

Data for the window overlap for window size $W=6$ and
different lattice sizes are shown in Fig.~\ref{fig:P60a_inset}. Even 
our largest systems can be
equilibrated, as indicated by the data dropping to to a very small value
($ < 10^{-3}$) at the longest times. Note that the value of $P_W(0)$ is not exactly zero even
when the system has fully equilibrated because of rare thermal fluctuations.
At short times the decay is roughly 
$t^{-1/2}$ as expected from coarsening~\cite{bray:02}, according to which
$\xi(t)$, the typical domain size, grows proportional to $t^{1/2}$.
However, in addition, a plateau
appears at intermediate times. We shall see that this plateau occurs because
in some runs, even when the correlation length has grown to the size of the
system, a single domain with straight walls persists for a much longer time.
Evidence for this is shown in the
inset to Fig.~\ref{fig:P60a_inset} which plots the probability of finding two
large clusters of oppositely oriented spins. This quantity has plateaus for the
same range of time as the data for $P_6(0)$ shown in the main part of the figure.

\begin{figure}
\begin{center}
\includegraphics[width=\columnwidth]{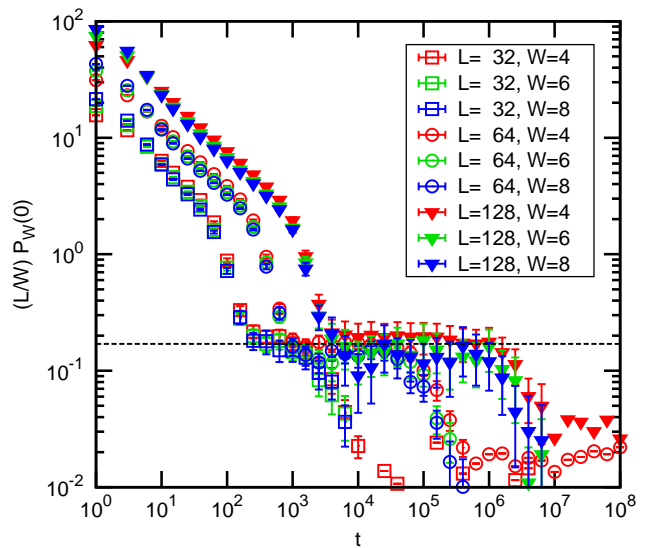}
\caption{(Color online)
Plot of $P_W(0)$ multiplied by the ratio of the system size $L$ to the window
size $W$ for
different values of $W$ and $L$ at $T=3.6 \simeq
0.80 T_c$ 
for the ferromagnet. On this plot the height of the plateau at intermediate
times seems to be independent of $L$ and $W$ showing that $P_W(0)$ itself is
proportional to $W/L$, which we interpret as the probability that a straight
domain wall goes through the window. The dashed horizontal line is a guide to
the eye.
\label{fig:PB0a}
}
\end{center}
\end{figure}

Figure \ref{fig:PB0a} plots data for different system sizes and window sizes,
and shows that the \textit{height} of the plateau is proportional to $W/L$ which has a
straightforward interpretation as the probability that a straight domain wall
passes through the window.

We find that the \textit{time} at the \textit{beginning}
of the plateau varies as $L^2$, which
is expected since it
is the time for the coherence length to grow to the system size according to
the coarsening picture in which $\xi(t) \propto t^{1/2}$. The time 
at the \textit{end} of the plateau grows more rapidly and we find empirically
it is roughly proportional to $L^{4.8}$.
We presume that this is the time scale needed
for a random walk of the (straight) domain walls to cause the domains to meet and
form one big domain. S.\ Redner (private communication) has argued that this
exponent is exactly four, and our data is consistent with this value.

The rich dynamics of three-dimensional Ising ferromagnets
after a quench have been studied in great detail, see e.g.~\cite{redner:11},
at very low and
zero temperature. By contrast, our results are for a much higher temperature,
though still below $T_c$. Based on the preliminary findings presented 
here, we feel it would be interesting to study this region
in more detail in the future.

\section{Conclusions}
\label{sec:concl}

We have shown that the non-equilibrium window overlap distribution
of a spin glass following a
quench to below $T_c$ can be well characterized by the ratio of the dynamic
coherence length $\xi(t)$ to the window size $W$. For a fixed $W$
the distribution tends to a well defined
limit at long times such that $\xi(t) \gg W$ but where $\xi(t)$ is still much
less than the system size $L$. This distribution depends strongly on $W$; for
example $P_W(0) \propto W^{\alpha/2}$ where $\alpha \simeq 0.44$.

However, if we can run the simulation for sufficiently long times that the
system globally equilibrates, i.e.~$\xi(t) \simeq L/2$, then there is a
change in behavior, which is abrupt when plotted on a logarithmic time scale,
see Fig.~\ref{fig:P_4_0_L},
such that $P_W(0)$ then only 
depends weakly on $W$ and is very similar to the value at zero overlap of
Parisi's global overlap distribution $P(q)$. Though a similar 
looking plateau was found for the not-quite fully 
equilibrated ferromagnet, characterized by the existence of domain walls,
it was qualitatively different since the height of the plateau depends on
the overall system size $L$ for the ferromagnet. By contrast, for
the spin glass, the data in Fig.~\ref{fig:P_4_0_L}, while admittedly not
\textit{fully} in the plateau region, does not show any dependence on $L$
until the final equilibrium is reached (the \textit{end} of the plateau).

The strong change in behavior for the spin glass when $\xi(t) \simeq L/2$
indicates that local spin correlations are sensitive to spin orientations, and
presumably also to the values of the interactions, at large distances.
According to the droplet theory, the local state of the system does
\textit{not} depend on the values of the interactions sufficiently far away. 
If the droplet theory is correct asymptotically,
the length-scale beyond which this independence occurs must be much larger
than the system sizes we have been able to equilibrate below $T_c$
(namely $L = 16$).

In addition, we have speculated on a possible connection between the
non-equilibrium dynamics discussed here and averages computed theoretically
using the ``metastate". For a future better understanding of 
this possible connection via numerical simulations
 a more intense use of powerful yet rather cheap devices as GPUs, like
in the present work \cite{manssen:14}, or the application of 
new algorithms like population annealing to spin glasses
\cite{machta2010,wang2014} might be useful.

\begin{acknowledgments}
MM and AKH thank Martin Weigel for interesting discussions.
We would like to thank Nick Read and David Huse for a most informative
correspondence, and Sid Redner for suggesting the exact value of the exponent
giving the length of the plateau in Fig.~\ref{fig:PB0a}.
The work of APY is supported in part by the National
Science Foundation under Grant
No.~DMR-1207036, and by a Research Award from the Alexander von Humboldt
Foundation. 

\end{acknowledgments}

\bibliography{refs}

\end{document}